\documentclass{moriond}

\usepackage{float}
\usepackage{graphicx} 
\usepackage{amsmath}
\usepackage{tabu}
\usepackage{comment}
\usepackage{enumerate}
\usepackage{amssymb}
\usepackage{caption}
\usepackage{tikz}
\usetikzlibrary{decorations.pathmorphing,decorations.pathreplacing,decorations.markings}

\def\Journal#1#2#3#4{{#1} {\bf #2}, #3 (#4)}

\tikzset{
  on each segment/.style={
    decorate,
    decoration={
      show path construction,
      moveto code={},
      lineto code={
        \path [#1]
        (\tikzinputsegmentfirst) -- (\tikzinputsegmentlast);
      },
      curveto code={
        \path [#1] (\tikzinputsegmentfirst)
        .. controls
        (\tikzinputsegmentsupporta) and (\tikzinputsegmentsupportb)
        ..
        (\tikzinputsegmentlast);
      },
      closepath code={
        \path [#1]
        (\tikzinputsegmentfirst) -- (\tikzinputsegmentlast);
      },
    },
  },
  mid arrow/.style={postaction={decorate,decoration={
        markings,
        mark=at position .55 with {\arrow[#1]{stealth}}
      }}},
      end arrow/.style={postaction={decorate,decoration={
        markings,
        mark=at position .9 with {\arrow[#1]{stealth}}
      }}},
      beg arrow/.style={postaction={decorate,decoration={
        markings,
        mark=at position .2 with {\arrow[#1]{stealth}}
      }}},
}

\newcommand{\Photo}{}

\newcommand{\amu}{(g-2)_\mu}
\newcommand{\C}[2]{\mathcal{C}_{\substack{#1 \\ #2}}}

\title{Is the $(g-2)_\mu$ anomaly a threat to Lepton Flavor Conservation?}

\author{Julie~Pag\`es}
\address{Physik-Institut, Universit\"at Z\"urich, CH-8057 Z\"urich, Switzerland}

\begin{document}
\vspace*{2cm}

\maketitle

\abstracts{
Although the $(g-2)_\mu$ anomaly can be explained by New Physics (NP) involving only muons, a more general flavor structure is usually expected for NP operators in the Standard Model (SM)  Effective Field Theory (SMEFT).
In particular, if one wants to provide a combined explanation of several beyond the SM effects, like Lepton Flavor Universality (LFU) Violation, as indicated by the B anomalies, then a strong alignment of the NP operators in flavor space is required to satisfy the bounds from observables featuring Lepton Flavor Violation (LFV), like $\mu \to e \gamma$.
We derived the tight bound of $10^{-5}$ on the flavor angle in the dipole operator in the charged-lepton mass basis in the SMEFT. 
We found that misalignment in several operators at high-scale could spoil the alignment at low-scale, through the Renormalization Group Evolution (RGE) of the SMEFT. In particular, it imposes constraints on some 4-fermions operators.
We explored dynamical mechanisms as well as flavor symmetries to explain this flavor alignment, and illustrated the difficulty to reach it in an explicit NP model.
If the $\amu$ anomaly is confirmed, the only natural explanation seems to lie in the individual lepton number conservation. If this accidental symmetry of the SM is also present in higher-order terms in the SMEFT, we are led to conclude that quark and lepton sectors behave quite differently beyond the SM. This proceeding is based on arXiv:2111.13724.
}

\section{Introduction}

The SM of particle physics exhibits an approximate accidental symmetry in the lepton sector, \textit{i.e.} $U(3)_\ell \times U(3)_e$, broken down to $U(1)_{L_e} \times U(1)_{L_\mu} \times U(1)_{L_\tau}$ by the charged leptons Yukawa coupling. 
This has two important consequences on the expected behavior of elementary particles with respect to leptons. 
First, we expect LFU, meaning each gauge boson should couple with the same strength to each family of leptons.
Secondly, LFV is forbidden. We know LFV must happen in nature because of neutrino oscillations, however only a soft breaking proportional to the neutrino masses, is needed, and in this case, individual lepton numbers remain a very good approximate symmetry.

Recent anomalies in observables involving leptons, and in particular muons, will hopefully help us shed light on the true behavior of lepton flavors. 
The muon anomalous magnetic moment, $a_\mu = (g_\mu -2)/2$, seems to deviate from the SM expectation \cite{2006.04822}. The recent measurement by the E989 experiment at FNAL \cite{2104.03281}, combined with the previous BNL result \cite{hep-ex/0602035}:
\begin{equation}
		\Delta a_\mu = a_\mu^{\rm Exp} - a_\mu^{\rm SM} = (251 \pm 59) \times 10^{-11}\, ,
\end{equation}
indicates a $4.2 \sigma$ discrepancy.
This result, taken alone, does not indicate violation of any symmetries mentioned above. However, the flavor anomalies in neutral and charged currents semi-leptonic B-decays~\cite{2111.13724} challenge LFU. 
Since there have been attempts to explain all of them together (see for example refs.~\cite{2104.05730,2110.07454,2104.03228}), one might ask what flavor structure a combined explanation requires. 
Explanations for LFU violation can lead to LFV and experimental constraints on charged LFV are already tight (as set by MEG~\cite{1605.05081})
\begin{equation}
	\mathcal{B} (\mu^+ \to e^+ \gamma)< 4.2 \times 10^{-13} ~[90 \% \rm C.L.]
\end{equation}
These indications imply a strong alignment of New Physics (NP) in flavor space which will be described and quantified in the context of the SMEFT and its RGE in the following.

\section{$\amu$ and Lepton Flavor Violation in the SMEFT RGE}

\subsection{Connection between $\amu$ and LFV}

The SMEFT allows to study the effects of heavy NP, above the electroweak scale, on low-energy observables without explicitly specifying the new heavy degrees of freedom. We can write modifications of the SM Lagrangian as higher dimensional operators
\begin{equation} \label{SMEFTL}
	\mathcal{L}_{\rm SMEFT} = \mathcal{L}_{\rm SM} + \sum_i \,\mathcal{C}_i \,\mathcal{O}_i\,
\end{equation}
where each Wilson coefficient $\mathcal{C}_i$ has dimension $1/\Lambda^{[\mathcal{O}_i]-4}$ where $\Lambda$ is the NP scale and $[\mathcal{O}_i]$ is the dimension of the operator $\mathcal{O}_i$. In order to account for the deviation in $\amu$, one must add the dipole operator
\begin{equation}
	\mathcal{O}_{\substack{e\gamma \\ rs}} = \frac{v}{\sqrt 2} \bar e_{Lr} \sigma^{\mu \nu} e_{Rs} F_{\mu\nu}\,.
\end{equation}
This same operator can also mediate the LFV processes, \textit{e.g.} $\mu \to e \gamma$. The contributions to the two processes were calculated at tree-level and used to estimate the size, or put a bound, on the corresponding Wilson coefficients:
\begin{align} \label{coeffsize}
	\Delta a_\mu &= \frac{4 m_\mu v}{e\sqrt 2 } ~\text{Re} ~\C{e\gamma}{22}' 
	\quad &\mathcal{B} (\mu^+ \to e^+ \gamma) &= \frac{m_\mu^3 v^2}{8\pi \Gamma_\mu } \left( |\C{e\gamma}{12}'|^2 + |\C{e\gamma}{21}'|^2 \right) \\ \nonumber
	&\Rightarrow ~\text{Re} ~\C{e\gamma}{22}' = 1 \times 10^{-5} \text{ TeV}^{-2}\,,
	\quad &&\Rightarrow ~  |\C{e\gamma}{12}'|< 2 \times 10^{-10} \text{ TeV}^{-2}\,,
\end{align}
where the prime denotes coefficients in the charged lepton mass basis.
We see that the two requirements above lead to a tight constraint on the off-diagonal elements in the $2\times 2$ light lepton sector: 
\begin{equation} \label{eps12}
	\left|\epsilon_{12}^{L(R)}\right| \equiv \left |\frac{\C{e\gamma}{12(21)}'}{\C{e\gamma}{22}'}\right| < 2 \times 10^{-5}
\end{equation}
Similar bounds and considerations can also be applied to the $\mu-\tau$ sector (see ref.~\cite{2111.13724}).

\subsection{RGE of the SMEFT and mass diagonalization} \label{RGESMEFT}

When NP is integrated out at a scale $\mu_H$, it will generate a certain set of operators and associated Wilson coefficients $C_i$ as in eq.~\ref{SMEFTL}. The coefficients then need to be run down to the low scale $\mu_L$ using the RGE equations:
\begin{equation} \label{opmix}
	\mu \frac{d}{d\mu}C_i = \frac{1}{16\pi^2} \beta_i \quad \text{where } \beta_i = \sum_{j}\gamma_{ij} \,C_j 
\end{equation}
with solution $\displaystyle C_i (\mu_L) = C_i (\mu_H) - \hat L \,\beta_i$ with $\hat L = \frac{1}{16 \pi^2} \log \left( \frac{\mu_H}{\mu_L}\right)$. Notice, by looking at the definition of $\beta_i$, that the RGE can mix different operators in the SMEFT, possibly with some not directly generated by the NP at the scale $\mu_H$.
Once the coefficients are expressed at $\mu_L$, one can go to the charged lepton mass basis by applying the rotation
\begin{equation} 
	\Theta_{L(R)}^\mathcal{Y} = - \left.\frac{[\mathcal{Y}_e]_{12(21)}}{[\mathcal{Y}_e]_{22}}\right|_{\mu_L}
\end{equation}
giving the dipole coefficients
\begin{equation} \label{rot}
	\C{e\gamma}{12}' (\mu_L) = \C{e\gamma}{12}(\mu_L) + \Theta_L^{\mathcal Y} ~\C{e\gamma}{22}(\mu_L)\,, \qquad\quad \C{e\gamma}{22}'(\mu_L) \approx \C{e\gamma}{22}(\mu_L) \,.
\end{equation}
Notice that a sizable $22$ entry of the dipole coefficients is necessary to explain the $\amu$ anomaly, as seen in eq.~\ref{coeffsize}, and will therefore bring a sizable contribution to the $12$ entry in the mass basis.

In the following table, we identified all operators in the SMEFT that can contribute to spoiling the constraint on the $\epsilon_{12}^L$ ratio in eq.~\ref{eps12}, either by operators mixing through RGE, eq.~\ref{opmix}, or by rotation to the mass basis, eq.~\ref{rot}, or both. We used the SMEFT RGE defined in ref.~\cite{1308.2627}.

\begin{center}
$\begin{tabu}{|l|l|}
\hline &\\[-8pt]
 \text{Broken phase}&\centering \text{Unbroken phase} \\[0.4pt]
 \hline &\\[-8pt]
  \mathcal{O}_{\substack{e\gamma\\rs}} = \dfrac{v}{\sqrt 2} \bar e_{Lr} \sigma^{\mu \nu} e_{Rs} F_{\mu\nu}
	&	
		O_{\substack{eB\\rs}} = \bar \ell_{Lr} \sigma^{\mu \nu} e_{Rs} \, H B_{\mu\nu}
     \\[2pt]
   {\color{gray} \mathcal{O}_{\substack{eZ\\rs}} = \dfrac{v}{\sqrt 2} \bar e_{Lr} \sigma^{\mu \nu} e_{Rs} Z_{\mu\nu}}
	&
	O_{\substack{eW\\rs}} = \bar \ell_{Lr} \sigma^{\mu \nu} e_{Rs}  \, \tau^I H \,W_{\mu\nu}^I
     \\[2pt]
     \hline &\\[-8pt]
    \mathcal{O}_{\substack{\mathcal{Y}_e\\rs}} = \dfrac{v}{\sqrt 2} \bar e_{Lr} e_{Rs} 
	&
		O_{\substack{Y_e\\rs}} =  \bar \ell_{Lr} e_{Rs} \, H
     \\[2pt]
    {\color{gray}\mathcal{O}_{\substack{\mathcal{Y}_{he}\\rs}} = \dfrac{h}{\sqrt 2} \bar e_{Lr} e_{Rs} }
	&
		O_{\substack{eH\\rs}} = \bar \ell_{Lr} e_{Rs}\,  H (H^\dag H)
      \\[2pt]   
      \hline &\\[-8pt]
		{\color{gray} \mathcal{O}^{(3)}_{\substack{eeuu\\prst}} \,,~\mathcal{O}^{(3)}_{\substack{\nu edu\\prst}}}
	&
		O^{(3)}_{\substack{lequ\\prst}} = (\bar \ell_{Lp}^j \sigma^{\mu\nu} e_{Rr}) \epsilon_{jk}(\bar q_{Ls}^k \sigma_{\mu\nu} u_{Rt}) 
       \\[2pt]
       {\color{gray}\mathcal{O}^{(1)}_{\substack{eeuu\\prst}} \,, ~\mathcal{O}^{(1)}_{\substack{\nu edu\\prst}}}
	&
		O^{(1)}_{\substack{lequ\\prst}} = (\bar \ell_{Lp}^j  e_{Rr}) \epsilon_{jk}(\bar q_{Ls}^k  u_{Rt}) 
 	\\[2pt]
 	\hline
\end{tabu}$
\captionof{table}{Relevant operators for the analyses. For completeness, all operators have been included but only the ones in black will be used in the discussion. \label{tab:relop}}
\end{center}

The change of basis from the unbroken phase to the broken phase is obtained by
\begin{equation}
	\begin{pmatrix}
		\C{e\gamma}{} \\ \C{eZ}{}
	\end{pmatrix}
	=
	\begin{pmatrix}
		c_{\theta_W}  & -s_{\theta_W} \\ -s_{\theta_W} & -c_{\theta_W}
	\end{pmatrix}
	\begin{pmatrix}
		C_{eB} \\ C_{eW}
	\end{pmatrix}\,,
	\qquad
	\begin{pmatrix}
		\mathcal{Y}_e \\ \mathcal{Y}_{he}
	\end{pmatrix}
	=
	\begin{pmatrix}
		1  & -\frac12 \\ 1 & -\frac32
	\end{pmatrix}
	\begin{pmatrix}
		Y_e \\ v^2 C_{eH}
	\end{pmatrix}\,,
\end{equation}	
where $\theta_W$ is the weak mixing angle.
\subsection{Flavor alignment definition}
In the following,  we will make a few assumptions to isolate the leading contributions to $\epsilon_{12}^L$. We will neglect all contributions proportional to gauge coupling squared, the quartic of the Higgs and all the Yukawa couplings but the top. Furthermore, we will assume $C_{eH}$ and $Y_e$ to be flavor aligned at the high-scale.
The full RGE equations are presented in the paper ref.~\cite{2111.13724}. Here we summarize the main results. 
The 4-fermion operator $O_{lequ}^{(3)}$ mix directly with the dipole operator $\mathcal{O}_{e\gamma}$, while the 4-fermion operators $O_{lequ}^{(1)}$ mix with the Yukawa $\mathcal{Y}_e$, which after rotation to the mass basis can spoil $\epsilon_{12}^L$ bound.
From our initial assumptions, both 4-fermion operator are with the top quark only, \textit{i.e.} $s=t=3$.
Finally the flavor alignment master formula can be written as
\begin{equation} \label{epsalign}
	\epsilon_{12}^L \equiv \left. \frac{\C{e\gamma}{12}'}{\C{e\gamma}{22}'}\right|_{\mu_L} = \left(\theta_L^{e\gamma} -\theta_L^Y \right) + \left(\theta_L^{lequ^{(3)}} -\theta_L^{e\gamma} \right) \Delta_3 +  \left(\theta_L^{lequ^{(1)}} -\theta_L^Y \right) \Delta_1
\end{equation}
where $\Delta_3 = \dfrac{-16 \hat{L} e y_t}{\C{e\gamma}{22}(\mu_L)} C^{(3)}_{\substack{lequ \\ 2233}} (\mu_H)$ and $\Delta_1 = \dfrac{-6 \hat{L} y_t^3 v^2}{[\mathcal{Y}_e]_{22}(\mu_L)} C^{(1)}_{\substack{lequ \\ 2233}} (\mu_H)$  and we defined the flavor angles or flavor phases as 
\begin{equation}
	\theta_L^X = \left.\dfrac{C_{\substack{X\\12}}}{C_{\substack{X\\22}}}\right|_{\mu_H} \,.
\end{equation}
The full expression for $\epsilon_{12}^L$ have to satisfy the bound eq.~\ref{eps12}, implying alignment of the different flavor phases involved $\theta_L^Y$, $\theta_L^{e\gamma}$ ,$\theta_L^{lequ^{(3)}}$ and $\theta_L^{lequ^{(1)}}$.
In the next section, we will discuss how to reach this alignment, either dynamically or with the help of flavor symmetries.

\section{Flavor alignment mechanisms}

\subsection{Dynamical conditions}
We will discuss three scenarios to reach the flavor alignment required in the previous section.
\begin{enumerate}[A.]
	\item In the case where both the Yukawa coupling $Y_e$ and the dipole coefficient $C_{e\gamma}$ are \underline{not} generated at the high-scale $\mu_H$, this is equivalent to setting $\theta_L^{e\gamma}=\theta_L^Y=0$. Then the muon Yukawa coupling and the dipole operator gets generated at low-scale $\mu_L$ by $C_{lequ}^{(1)}$ and $C_{lequ}^{(3)}$, respectively, implying $\Delta_3=-\Delta_1=1$. Now, if additionally the same NP dynamics generate $C^{(1)}_{lequ}$ and $C^{(3)}_{lequ}$ at high-scale, then $\theta_L^{lequ^{(3)}}= \theta_L^{lequ^{(1)}}$ and we finally reach the required flavor alignment.
	\item In the case where the Yukawa coupling and the dipole coefficient are generated at the high-scale, then they should be generated by the same NP dynamics to obtain $\theta_L^{e\gamma}= \theta_L^Y$. A simple option at this point is that the 4-fermions operators in table~\ref{tab:relop} are not generated at high-scale, leading to $\Delta_3=\Delta_1=0$, which removes the need to align the remaining flavor phases.
	\item Finally, the third straightforward option is to generate all four operators with the same NP dynamics, leading to all the flavor phases being aligned at the high-scale, \textit{i.e.} $\theta_L^{e\gamma} = \theta_L^{Y} = \theta_L^{lequ^{(3)}}= \theta_L^{lequ^{(1)}} $.
\end{enumerate}
The three options are summarized in table~\ref{tab:scenarios}.
\begin{table}[H]
\begin{center}
	\begin{tabular}{|c|c|c|c|}
	\hline
		Scenario & A\hspace*{0.2cm} & \hspace*{1.2cm}B\hspace*{1.4cm}  & \hspace*{1.2cm}C\hspace*{1.4cm}  \\
		\hline &&&\\[-0.4cm]
		Constraints at $\mu_H$   & 
				$C_{e\gamma} = Y_e = 0$ \hspace*{0.2cm} & 
				$C_{lequ}^{(1)} = C_{lequ}^{(3)} = 0$ \hspace*{0.1cm} & 
				\hspace*{0.3cm} none \hspace*{0.6cm} \\[0.1cm]
		\hline&&&\\[-0.4cm]
		Alignment condition & 
		\hspace*{0.1cm} $\theta_L^{lequ^{(3)}}-~\theta_L^{lequ^{(1)}}$ \hspace*{0.1cm} &  
		$\theta_L^{e\gamma}-~\theta_L^{Y}$\hspace*{0.2cm} &  
		full expr. eq.~\ref{epsalign}
		\\[0.1cm]
		\hline
	\end{tabular}
	\caption{\label{tab:scenarios} Dynamical mechanisms for flavor alignment. The remaining alignment condition in the last row is simply the $\epsilon_{12}^L$ expression from eq.~\eqref{eps12} after applying the constraints at $\mu_H$ coming from our scenario assumption.}
	\end{center}
\end{table}
In all three cases, the flavor alignment seems rather unnatural and requires many non-trivial assumptions.

\subsection{Flavor symmetries}
We also explored if and how lepton flavor symmetries can help with the alignment of the flavor phases.
\begin{enumerate}[A.]
	\item $U(2)_\ell \times U(2)_e$ symmetry\\
This symmetry, as originally presented in ref.~\cite{1105.2296}, assumes that light families of leptons are massless and indistinguishable.
It needs to be minimally broken by two spurions $V_\ell \sim (\mathbf 2, \mathbf 1)$ and $\Delta_e \sim (\mathbf 2, \mathbf {\bar 2})$ such that the charged-lepton Yukawa coupling assumes the form
\begin{equation}
	Y_e= y_\tau \begin{pmatrix}
		\Delta_e & V_\ell \\0 & 1
	\end{pmatrix} 
\end{equation}
with
	$V_\ell = \begin{pmatrix}
		0 \\ \epsilon_\ell
	\end{pmatrix}$ and
	$\Delta_e = \begin{pmatrix}
		c_e &-s_e \\ s_e & c_e
	\end{pmatrix}\begin{pmatrix}
		\delta_e' &0 \\ 0 & \delta_e
	\end{pmatrix} $ with $|\delta_e'|\ll |\delta_e|\ll |\epsilon_\ell|\ll1$. 
	We expect $\epsilon_\ell \sim \mathcal{O}(10^{-1})$ from the B anomalies (see ref.~\cite{1909.02519}) and a natural expectation for the angle is $s_e \gtrsim \mathcal{O}\left(\sqrt{m_e/m_\mu}\right)$.
	Generically we can factorize the flavor structure of the semi-leptonic operators discussed in table~\ref{tab:relop} and express them in terms of the spurions (see ref.~\cite{2005.05366}), up to order $\mathcal{O}(V_\ell^2 \Delta_e)$, as
	\begin{equation}
		X_{\alpha\beta}^n \left(\bar \ell_\alpha \Gamma^n e_\beta \right) \eta^n 
		\quad \text{with}
		\quad
		X_{\alpha\beta}^n = a_n (\Delta_e)_{\alpha \beta} + b_n (V_\ell)_\alpha(V^\dagger_\ell)_\gamma (\Delta_e)_{\gamma \beta} 
	\end{equation}
	where $\Gamma^n$ is the Dirac structure of the fermion product and $\eta^n$ is the lepton-independent part of the operator $n $, while $a_n$ and $b_n$ are $\mathcal{O}(1)$ coefficients.
	The flavor phases in this framework then read
	\begin{equation}
		\theta_L^n \approx \frac{s_e}{c_e} \left(1-\frac{b_n}{a_n}\epsilon_\ell^2\right)
	\end{equation}
	Notice that the flavor phases are exactly aligned in the case of fully minimal $U(2)_\ell \times U(2)_e$ symmetry, \textit{i.e.} $b_n=0$. However, some misalignment can be reintroduced by the sub-leading spurionic contribution, \textit{e.g.}
	\begin{equation}
		|\theta_L^{e\gamma} - \theta_L^{Y}| = \left| \frac{s_e}{c_e}\right|  \epsilon_\ell^2 \left| \frac{b_Y}{a_Y}-\frac{b_{e\gamma}}{a_{e\gamma}} \right| \leq 2 \times 10^{-5}
	\end{equation}
	We therefore expect a mild tuning of $10^{-2}$ in the difference of the ratio of $\mathcal{O}(1)$ coefficients $a_n$ and $b_n$. 
	The $U(2)_\ell \times U(2)_e$ symmetry is therefore helping the flavor alignment but is not enough to completely protect from LFV in the 1-2 sector. 
	Note also that since $s_e$ is not fixed, one could also assume a smaller $s_e$ spurion from the start to satisfy the bound without the need to tune the $\mathcal{O}(1)$ coefficients. Extrapolating this direction by taking $s_e\to 0$ brings us to the next flavor symmetry we considered in this work.

	\item $U(1)_{L_e} \times U(1)_{L_\mu} \times U(1)_{L_\tau}$ symmetry\\
	In this symmetry, the charged-lepton Yukawa coupling and all flavor structures $X^n_{\alpha \beta}$ can only assume a fully diagonal form. Therefore, all flavor phases vanish and no alignment is needed. One can also use the anomaly free combination $U(1)_{L} \times U(1)_{L_e-L_\mu} \times U(1)_{L_\mu-L_\tau}$ where $U(1)_{L}$ is lepton number.
	In particular, any combination of $U(1)_{L_\mu}$ and $U(1)_{L_\tau}$ is enough to protect the $\mu-e$ mixing.
\end{enumerate}

\subsection{Example: an explicit NP model}
In this last part, we add two simple mediators to the SM Lagrangian which will generate at tree-level, the 4-fermions operators in table~\ref{tab:relop}, namely, a scalar leptoquark $S_1 \sim (\mathbf{\bar 3}, \mathbf{1})_{1/3}$ and a Higgs-like field $\Phi\sim (\mathbf{1}, \mathbf{2})_{1/2}$.

The full Lagrangian reads
\begin{align}
	\mathcal L ~=~ &\mathcal L_{\rm SM} + (D_\mu S_1)^\dag (D^\mu S_1) - M_{S_1}^2 S_1^\dag S_1 - \left[ \lambda^L_{i\alpha} (\bar q_i^c \epsilon \ell_\alpha) S_1 +  \lambda^R_{i\alpha} (\bar u_i^c e_\alpha) S_1 + \rm h.c. \right] \\
	&+ (D_\mu \Phi)^\dag (D^\mu \Phi) - M_\Phi^2 \Phi^\dag \Phi -\left[ \lambda^e_{\alpha \beta} (\bar \ell_\alpha e_\beta) \Phi +  \lambda^u_{ij} (\bar q_i u_j) \tilde \Phi + \rm h.c. \right]  \nonumber\,.
\end{align} 
The precise SMEFT matching coefficients after integrating out $S_1$ and $\Phi$ are presented in the paper ref.~\cite{2111.13724}. Here we discuss how to reach the flavor alignment in this specific model by illustrating the different contributions and describing what assumptions are necessary to align the flavor phases.-
The scalar leptoquark $S_1$ generates $C^{(1)}_{lequ}$ and $C^{(3)}_{lequ}$ at tree-level, at the same time via fierzing:
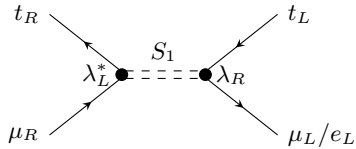
\begin{center}
\begin{tikzpicture}
\path[draw=black,postaction={on each segment={mid arrow=black}}]
(-1.5,-.8) node[left] {\footnotesize$\mu_R$} 
--  (-0.5,0) -- (-1.5,.8) node[left] {\footnotesize$t_R$}
;
\path[draw=black,postaction={on each segment={mid arrow=black}}]
(1.5,.8) node[right] {\footnotesize$t_L$}
--  (0.5,0) -- (1.5,-.8)  node[right] {\footnotesize$\mu_L/e_L$} 
;
\filldraw [black] (-0.55,0) circle (2.2pt) node[left]{\footnotesize$\lambda_L^*$};
\filldraw [black] (0.55,0) circle (2.2pt) node[right]{\footnotesize$\lambda_R$};
\node at (0,0.3) {\footnotesize$S_1$};
\draw[dashed] (-0.5,0.05) -- (0.5,0.05);
\draw[dashed] (-0.5,-0.05) -- (0.5,-0.05);
\end{tikzpicture}
\end{center}
as well as $C_{e\gamma}$ at the one-loop level:
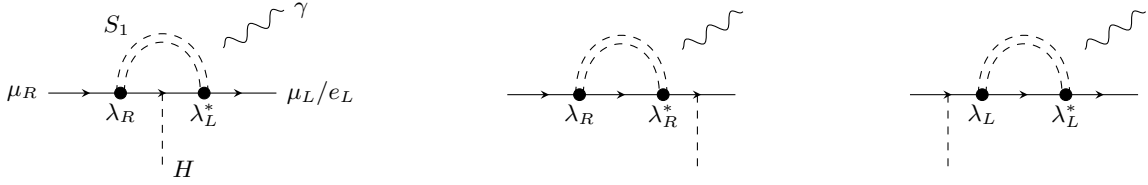
\begin{center}
\begin{minipage}{0.33\textwidth}
\begin{tikzpicture}
\path[draw=black,postaction={on each segment={mid arrow=black}}]
(-1.5,0) node[left] {\footnotesize$\mu_R$}
--  (-0.5,0) -- (0.5,0) -- (1.5,0) node[right] {\footnotesize$\mu_L/e_L$} 
;
\node at (-0.6,0.9) {\footnotesize$S_1$};
\filldraw [black] (-0.55,0) circle (2.2pt) node[below]{\footnotesize$\lambda_R$};
\filldraw [black] (0.55,0) circle (2.2pt) node[below]{\footnotesize$\lambda_L^*$};
\draw[dashed] ((-0.5,0) .. controls (-0.5,0.9)  and (0.5,0.9) .. (0.5,0);
\draw[dashed] ((-0.6,0) .. controls (-0.6,1.05)  and (0.6,1.05) .. (0.6,0);
\draw [-,decorate,decoration=snake] (0.8,0.6) -- (1.6,1.1) node[right] {\footnotesize$\gamma$};
\draw[dashed] (0,0) -- (0,-1) node[right] {\footnotesize$H$};
\end{tikzpicture}
\end{minipage}
\hfill
\begin{minipage}{0.25\textwidth}
\begin{tikzpicture}
\path[draw=black,postaction={on each segment={mid arrow=black}}]
(-1.5,0) --  (-0.5,0) -- (0.5,0) -- (1.5,0) ;
\filldraw [black] (-0.55,0) circle (2.2pt) node[below]{\footnotesize$\lambda_R$};
\filldraw [black] (0.55,0) circle (2.2pt) node[below]{\footnotesize$\lambda_R^*$};
\draw[dashed] ((-0.5,0) .. controls (-0.5,0.9)  and (0.5,0.9) .. (0.5,0);
\draw[dashed] ((-0.6,0) .. controls (-0.6,1.05)  and (0.6,1.05) .. (0.6,0);
\draw [-,decorate,decoration=snake] (0.8,0.6) -- (1.6,1.1) ;
\draw[dashed] (1,0) -- (1,-1) ;
\end{tikzpicture}
\end{minipage}\hfill
\begin{minipage}{0.25\textwidth}
\begin{tikzpicture}
\path[draw=black,postaction={on each segment={mid arrow=black}}]
(-1.5,0) --  (-0.5,0) -- (0.5,0) -- (1.5,0);
\filldraw [black] (-0.55,0) circle (2.2pt) node[below]{\footnotesize$\lambda_L$};
\filldraw [black] (0.55,0) circle (2.2pt) node[below]{\footnotesize$\lambda_L^*$};
\draw[dashed] ((-0.5,0) .. controls (-0.5,0.9)  and (0.5,0.9) .. (0.5,0);
\draw[dashed] ((-0.6,0) .. controls (-0.6,1.05)  and (0.6,1.05) .. (0.6,0);
\draw [-,decorate,decoration=snake] (0.8,0.6) -- (1.6,1.1) ;
\draw[dashed] (-1,0) -- (-1,-1);
\end{tikzpicture}
\end{minipage} 
\end{center}
while the Higgs-like $\Phi$ only generates $C^{(1)}_{lequ}$:
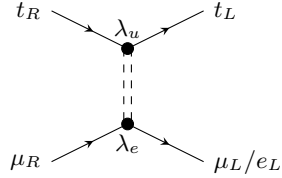
\begin{center}
\begin{tikzpicture}
\path[draw=black,postaction={on each segment={mid arrow=black}}]
(-1,-1) node[left] {\footnotesize$\mu_R$} 
--  (0,-0.5) -- (1,-1) node[right] {\footnotesize$\mu_L/e_L$}
;
\path[draw=black,postaction={on each segment={mid arrow=black}}]
(-1,1) node[left] {\footnotesize$t_R$} 
--  (0,0.5) -- (1,1) node[right] {\footnotesize$t_L$}
;
;
\filldraw [black] (0,0.5) circle (2.2pt) node[above]{\footnotesize$\lambda_u$};
\filldraw [black] (0,-0.5) circle (2.2pt) node[below]{\footnotesize$\lambda_e$};
\draw[dashed] (0.05,0.5) -- (0.05,-0.5);
\draw[dashed] (-0.05,0.5) -- (-0.05,-0.5);
\end{tikzpicture}
\end{center}

\begin{itemize}
	\item The first observation is that in the decoupling limit $M_\Phi \gg M_{S_1}$, only the first Feynman diagram contribute to the 4-fermion operators. Since fierzing preserves the flavor alignment of the original operator (with quark-lepton currents), we obtain $\theta_L^{lequ^{(3)}}=\theta_L^{lequ^{(1)}}$.
	\item The second assumption we can make is that the scalar leptoquark couples dominantly to left-handed tops. In that case, we can neglect the middle diagram in the loop-generated dipole and only the top runs inside the loop. We found with this that we obtain the alignment $\theta_L^{e\gamma}= \theta_L^{lequ^{(3)}}$.
	\item Finally, the simplest assumption we can make for the Yukawa coupling is that of a $U(2)$ flavor symmetry. Unfortunately, since the leading $U(2)$ spurion is not generated by integrating out the leptoquark, the misalignment we are left with is $\theta_L^{Y}-\theta_L^{e\gamma}\approx s_e$ and it can only satisfy the LFV bounds by requiring the unnatural condition $s_e \lesssim 10^{-5}$.
\end{itemize}
More generally, in this model we observed a tension in trying to align $\theta_L^{e\gamma}$ with $\theta_L^Y$ and, at the same time, with $\theta_L^{lequ^{(3)}}$.

\section*{Acknowledgments}
I would like to thank the Organizers of Les Rencontres de Physique de la Vall\'ee d'Aoste for the invitation, as well as my supervisor Gino Isidori and my colleague Felix Wilsch for the collaboration on this work.


\end{document}